\documentclass{chi-ext}

\copyrightinfo{Permission to make digital or hard copies of part or all of this work for personal or classroom use is granted without fee provided that copies are not made or distributed for profit or commercial advantage and that copies bear this notice and the full citation on the first page. Copyrights for third-party components of this work must be honored. For all other uses, contact the owner/author(s). Copyright is held by the author/owner(s). CHI'15 Workshop 05 "Collaborating with Intelligent Machines: Interfaces for Creative Sound", April 18, 2015, Seoul, Republic of Korea.}

\title{Machine Intelligence, New Interfaces, and the Art of the Soluble}

\numberofauthors{1}
\author{
  	\textbf{Michael J. Lyons}\\
  	\affaddr{Ritsumeikan, University}\\
  	\affaddr{Kyoto, Japan}\\
  	\email{michael.lyons@gmail.com}
  }

\def\plaintitle{Machine Intelligence, New Interfaces, and the Art of the Soluble}
\def\plainauthor{Michael J. Lyons}
\def\plainkeywords{Sound; music; art; interface;  musical interface}
\def\plaingeneralterms{Design; Experimentation; Human Factors; Performance}

\hypersetup{
  pdftitle={\plaintitle},
  pdfauthor={\plainauthor},  
  pdfkeywords={\plainkeywords},
  pdfsubject={\plaingeneralterms},
}

\usepackage{graphicx}   
\usepackage{balance}    
\usepackage{bibspacing} 

\begin{document}

\maketitle

\begin{abstract}
Position: (1) Partial solutions to machine intelligence can lead to systems which may be useful creating interesting and expressive musical works. (2) An appropriate general goal for this field is augmenting human expression. (3) The study of the aesthetics of human augmentation in musical performance is in its infancy.
\end{abstract}




\section{Background}
My interest in machine intelligence dates to the late 1980s and 1990s. Approaches to simulating intelligence shifted to more embodied approaches \cite{brooks1991intelligence, mead1989analog} which appealed to those with a background in natural science. A strong motivation for the work in analog computation, neural networks, and embodied robotics was the hunch that systems making parallel use of many more physical degrees of freedom than the usual von Neumann architectures could lead to powerful new approaches to computation. The extreme form of this hunch found expression in intensified research into quantum computing. In the mid-1990s it became increasingly clear that these approaches were also not living up to their initial promise of new solutions to hard problems in artificial intelligence. 
	Around this time, there was growing interest in combining machine intelligence approaches with human computer interfaces. This turned out to be an example of Ôthe art of the solubleÕ \cite{bartneck2007hci} - human-machine systems have proven to be a more realistic and practically useful goal than some of the difficult (albeit very interesting) research problems studied by both classical AI and physical computation researchers. Many human interface systems that have since been incorporated into consumer products (e.g. gesture recognizing touch surfaces) are commonplace examples. Of key importance is that these technologies augment or support human intelligence rather than attempt to function in an independently autonomous fashion. 

\section{Human-Computer Interaction and Musical Expression}
Just as research in machine intelligence and human-computer interaction were converging, I and some of my colleagues at the Advanced Telecommunications Research Labs in Kyoto became interested in musical applications of new human-computer interfaces, eventually leading us to organize a workshop on the topic at CHI 2001 \cite{NIME01}. Our interests resonated with other researchers who were attracted to participate and the annual NIME conference was born. The interests of the musical HCI community are diverse (and indeed are more extensive than NIME), but it has remained true that a core interest of this community is the integration of intelligent machine technology with human musical activity. However, it seems possible, if not likely, that there are as many understandings (some explicit, some implicit) of what constitutes intelligent systems as there are researchers participating in NIME.

\section{Intelligence and Music Technology}
Most explicitly recognizable is the direct application of machine learning methods in the design of musical interfaces \cite{fiebrink2010wekinator}. Example-based learning methods are useful in the design of complex mappings between control and synthesis parameters. The methods have been incorporated into systems that are sufficiently user-friendly to be useful for those with limited technical knowledge of machine learning. 

Some of the most convincing work on intelligent musical interfaces to date \cite{bevilacqua2013mo} combines cognitive studies of timbre, gestural action, with machine learning. This research project is perhaps the most advanced in leveraging an understanding of human perception/action  and machine learning methods to arrive at intuitively usable technology.

New musical interface research has also made broad use of techniques from other areas of computer science research such as computer vision to give just one example \cite{lyons2001facing, lyons2004facial}. Often, the incorporation of technology into interactive human perception/action loop renders partial solutions to difficult AI problems practically usable, even under the real-world, real-time demands of musical performance.

Likewise, in successful musical interface research, ÔintelligenceÕ takes the form of well chosen or designed affordances and mappings  \cite{fels2009creating}. This again is an embodied form of intelligence that takes into account and effectively relies on the human perception and action. 


\begin{figure}
  \centering
  \includegraphics[width=.95\linewidth]{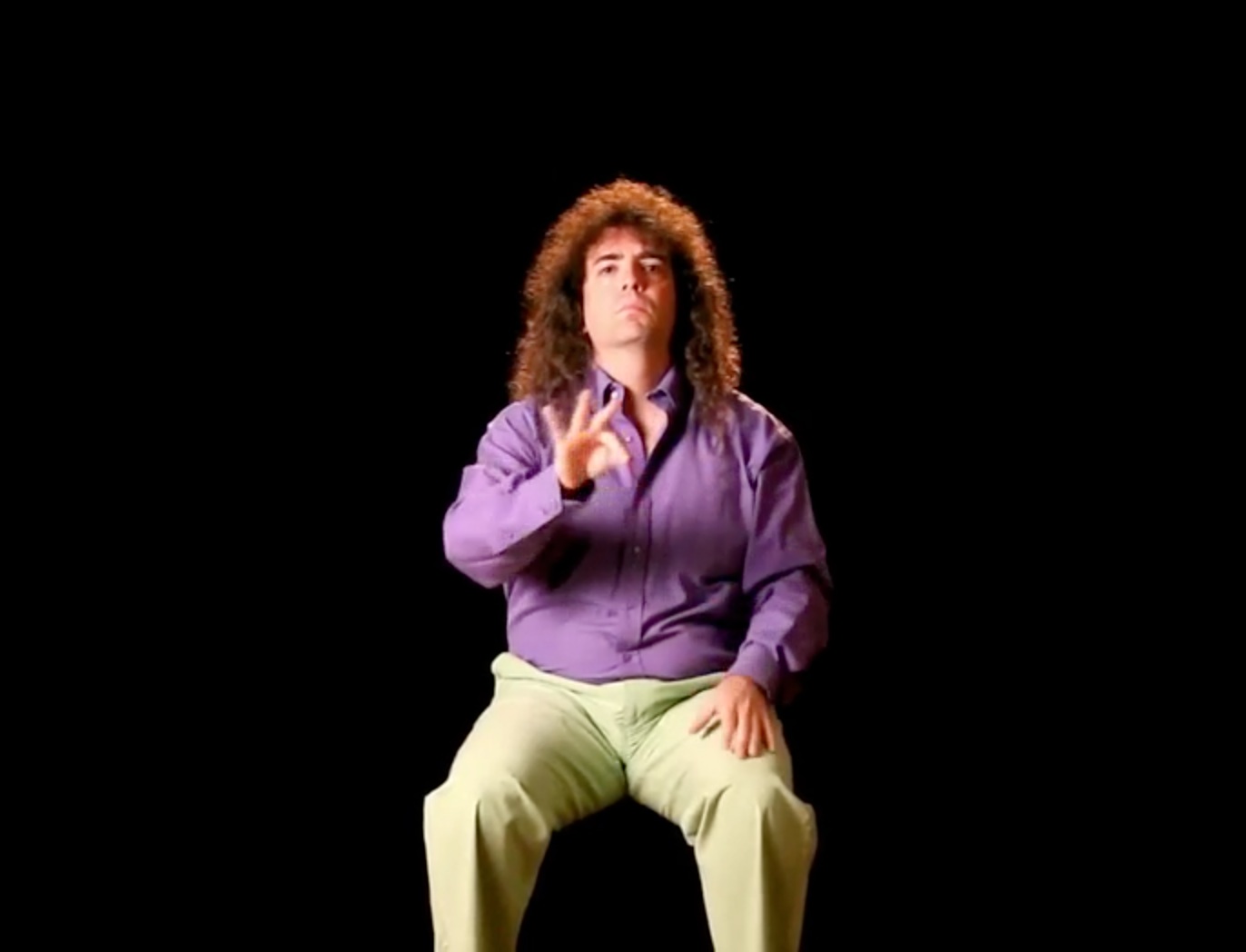}
  \caption{Mark Applebaum performing \textit{Aphasia}.}
  \label{fig:aphasia}
\end{figure}

\begin{figure}
  \begin{center}
  \includegraphics[width=.95\linewidth]{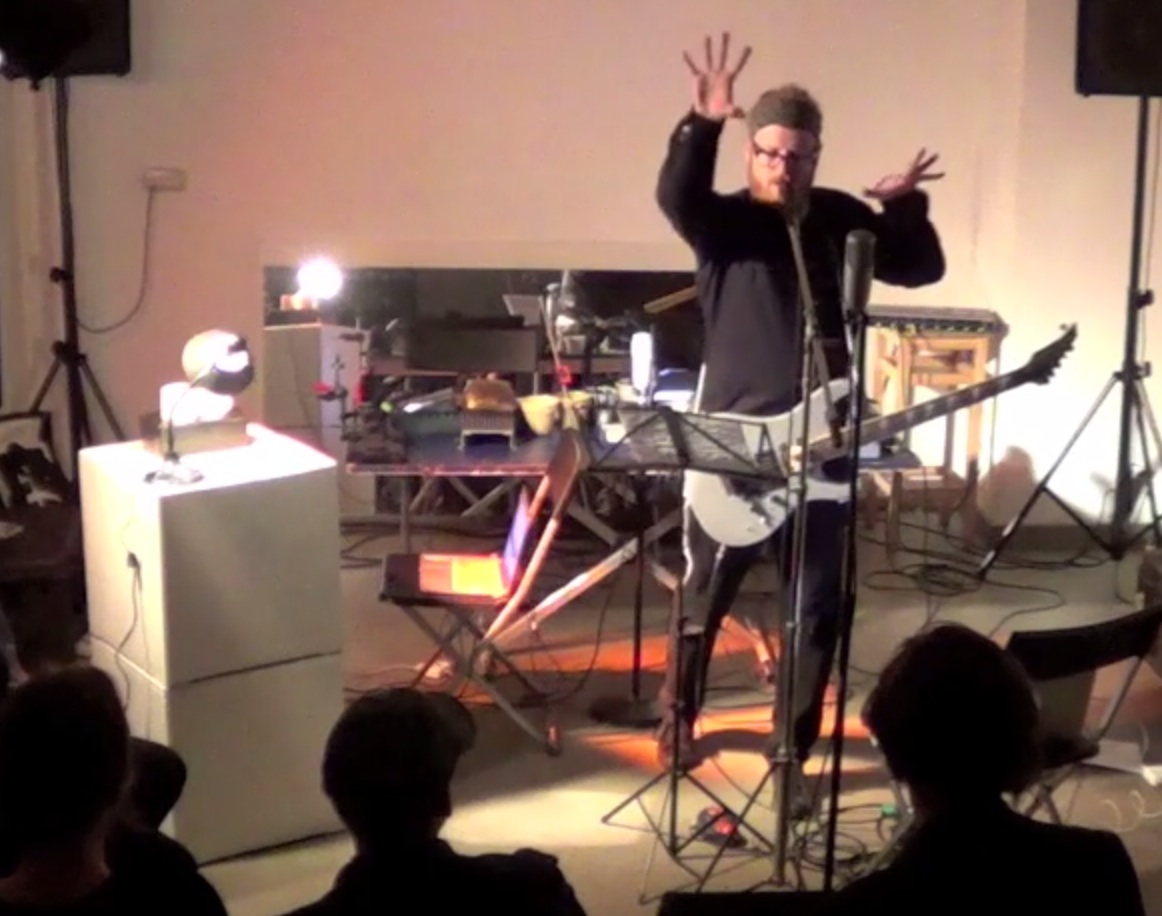}
  \caption{Dennis Sullivan performing Alexander Schubert`s \textit{Your Fox`s a Dirty Gold}.}
  \label{fig:your_fox}
  \end{center}  
\end{figure}

\section{Conclusion: Machine Intelligence and Performance Aesthetics}
Before concluding, I would like to introduce two recent and intriguing musical compositions. 

One is Mark Applebaum`s \textit{Aphasia}. In this piece, Applebaum performs a series of hand gestures synchronized to pre-recorded sound (processed vocal samples). Applebaum`s performance and versions by many other performers may be found on YouTube and are very worthwhile. I first experienced Applebaum`s own performance live as part of the concert program of NIME 12. I had not read the program notes and for the first minute was shocked to think that Applebaum might be controlling the sound in real-time with his complex and idiosyncratic gestures. Indeed, this piece is a kind of music technology ÔWizard-of-OzÕ experiment and could possibly serve in the future as a benchmark (the ÔApplebaum TestÕ?) for gesture recognition technology.

In a similar vein is Alexander Schubert`s \textit{Your Fox`s a Dirty Gold}: a work for solo performer with voice, motion sensors, electric guitar, and live electronics. Notable performances by Frauke Aulbert and Dennis Sullivan may be found on YouTube and Vimeo, respectively. This piece does involve gesturally controlled audio effects but the exact nature of the gestural mapping is not obvious. The theatricality of the performance plays a role in the success of the piece, as indeed does the tension caused by the semi-opaque gesture-to-sound mapping. 

Are \textit{Aphasia} and \textit{Your Fox`s a Dirty Gold} the kind of composition we might expect should real-time gesture recognition systems be perfected?  Would not these works be less interesting, humorous, and theatrical if such technology did actually exist? If so, what is the aim of developing intelligent musical performance technology? It is clear that the aesthetics of performance mediated by intelligent interface technology is a fascinating  and unexplored topic.

From the 20th century onwards experimental music often ingeniously involves chance operations, indeterminacy, silence, discarded instruments, hacked electronic devices, subverted tape machines and record players, the sounds of nature, the city, and every life. Such work is often more compelling than anything that has yet arisen from an analytical, technically driven approach. But there will also be new technologically mediated approaches to augmenting human expression. Music lives in the realm of shared subjective experience and it is open to invention.

\balance
\bibliographystyle{acm-sigchi}
\bibliography{lyons_chi15_workshop}

\begin{thebibliography}{1}

\bibitem{bartneck2007hci}
Bartneck, C., and Lyons, M.~J.
\newblock Hci and the face: Towards an art of the soluble.
\newblock In {\em Human-Computer Interaction. Interaction Design and
  Usability}. Springer, 2007, 20--29.

\bibitem{bevilacqua2013mo}
Bevilacqua, F., Schnell, N., Rasamimanana, N., Bloit, J., Flety, E., Caramiaux,
  B., Fran{\c{c}}oise, J., and Boyer, E.
\newblock De-mo: designing action-sound relationships with the mo interfaces.
\newblock In {\em CHI'13 Extended Abstracts on Human Factors in Computing
  Systems}, ACM (2013), 2907--2910.

\bibitem{brooks1991intelligence}
Brooks, R.~A.
\newblock Intelligence without representation.
\newblock {\em Artificial intelligence 47}, 1 (1991), 139--159.

\bibitem{fels2009creating}
Fels, S., and Lyons, M.
\newblock Creating new interfaces for musical expression: introduction to nime.
\newblock In {\em ACM SIGGRAPH 2009 Courses}, ACM (2009), 11.

\bibitem{fiebrink2010wekinator}
Fiebrink, R., and Cook, P.~R.
\newblock The wekinator: a system for real-time, interactive machine learning
  in music.
\newblock In {\em Proceedings of The Eleventh International Society for Music
  Information Retrieval Conference (ISMIR 2010). Utrecht} (2010).

\bibitem{lyons2004facial}
Lyons, M.~J.
\newblock Facial gesture interfaces for expression and communication.
\newblock In {\em Systems, Man and Cybernetics, 2004 IEEE International
  Conference on}, vol.~1, IEEE (2004), 598--603.

\bibitem{lyons2001facing}
Lyons, M.~J., and Tetsutani, N.
\newblock Facing the music: a facial action controlled musical interface.
\newblock In {\em CHI'01 extended abstracts on Human factors in computing
  systems}, ACM (2001), 309--310.

\bibitem{mead1989analog}
Mead, C., and Ismail, M.
\newblock {\em Analog VLSI implementation of neural systems}.
\newblock Springer Science \& Business Media, 1989.

\bibitem{NIME01}
Poupyrev, I., Lyons, M., Fels, S., and Blaine, T.
\newblock New interfaces for musical expression.
\newblock In {\em Extended Abstracts CHI'01}, ACM Press (2001), 491--492.

\end{thebibliography}

\end{document}